\documentclass{emulateapj}
\usepackage{float}
\usepackage{amsmath}
\usepackage{epsfig,floatflt}

\usepackage{graphicx}
\usepackage[]{subfigure}
\usepackage{color}
\definecolor{Blue}{rgb}{0.1,0.1,1.0} 
\definecolor{Magenta}{rgb}{1.0,0.1,0.5} 
\definecolor{LRed}{rgb}{0.8,0.0,0.0}

\newcommand{\nc}{\newcommand}

\nc{\be}[1]{\begin{equation}\mbox{$\label{#1}$}}
\nc{\bea}[1]{\begin{eqnarray} \mbox{$\label{#1}$}}
\nc{\Section}[2]{\section{#2}\label{#1}}
\nc{\Bibitem}[1]{\bibitem{#1}}
\nc{\Label}[1]{\label{#1}}

\nc{\eea}{\end{eqnarray}}
\nc{\ee}{\end{equation}}

\nc{\jcap}{JCAP}
\nc{\bdm}{\begin{displaymath}}
\nc{\edm}{\end{displaymath}}
\nc{\dpsty}{\displaystyle}
\nc{\bc}{\begin{center}}
\nc{\ec}{\end{center}}
\nc{\ea}{\end{array}}
\nc{\bab}{\begin{abstract}}
\nc{\eab}{\end{abstract}}
\nc{\btab}{\begin{tabular}}
\nc{\etab}{\end{tabular}}
\nc{\bit}{\begin{itemize}}
\nc{\eit}{\end{itemize}}
\nc{\ben}{\begin{enumerate}}
\nc{\een}{\end{enumerate}}
\nc{\bfig}{\begin{figure}}
\nc{\efig}{\end{figure}}

\nc{\arreq}{&\!=\!&}
\nc{\arrmi}{&\!-\!&}
\nc{\arrpl}{&\!+\!&}
\nc{\arrap}{&\!\!\!\approx\!\!\!&}
\nc{\non}{\nonumber}

\def\lsim{\; \raise0.3ex\hbox{$<$\kern-0.75em
      \raise-1.1ex\hbox{$\sim$}}\; }
\def\gsim{\; \raise0.3ex\hbox{$>$\kern-0.75em
      \raise-1.1ex\hbox{$\sim$}}\; }

\nc{\DOT}{\hspace{-0.08in}{\bf .}\hspace{0.1in}}
\nc{\Laada}{\hbox {$\sqcap$ \kern -1em $\sqcup$}}
\nc\loota{{\scriptstyle\sqcap\kern-0.55em\hbox{$\scriptstyle\sqcup$}}}
\nc\Loota{{\sqcap\kern-0.65em\hbox{$\sqcup$}}}
\nc\laada{\Loota}
\nc{\qed}{\hskip 3em \hbox{\BOX} \vskip 2ex}

\nc{\real}{{\rm I \! R}}
\nc{\Z}{{\sf Z \!\!\! Z}}
\nc{\complex}{{\rm C\!\!\! {\sf I}\,\,}}
\def\bigid{\leavevmode\hbox{\small1\kern-3.8pt\normalsize1}}
\def\id{\leavevmode\hbox{\small1\kern-3.3pt\normalsize1}}
\nc{\slask}{\!\!\!/}
\nc{\bis}{{\prime\prime}}
\nc{\pa}{\partial}
\nc{\ra}{\rangle}
\nc{\goto}{\rightarrow}
\nc{\swap}{\leftrightarrow}

\nc{\EE}[1]{ \mbox{$\cdot10^{#1}$} }
\nc{\abs}[1]{\left|#1\right|}
\nc{\at}[2]{\left.#1\right|_{#2}}
\nc{\norm}[1]{\|#1\|}
\nc{\abscut}[2]{\Abs{#1}_{\scriptscriptstyle#2}}
\nc{\vek}[1]{{\rm\bf #1}}
\nc{\integral}[2]{\int\limits_{#1}^{#2}}
\nc{\inv}[1]{\frac{1}{#1}}
\nc{\dd}[2]{{{\partial #1}\over{\partial #2}}}
\nc{\ddd}[2]{{{{\partial}^2 #1}\over{\partial {#2}^2}}}
\nc{\dddd}[3]{{{{\partial}^2 #1}\over
    {\partial #2 \partial #3}}}
\nc{\dder}[2]{{{d #1}\over{d #2}}}
\nc{\ddder}[2]{{{d^2 #1}\over{d {#2}^2}}}
\nc{\dddder}[3]{{d^2 #1}\over
    {d #2 d #3}}
\nc{\dx}[1]{d\,^{#1}x}
\nc{\dy}[1]{d\,^{#1}y}
\nc{\dz}[1]{d\,^{#1}z}
\nc{\dl}[1]{\frac{d\,^{#1}l}{(2\pi)^{#1}}}
\nc{\dk}[1]{\frac{d\,^{#1}k}{(2\pi)^{#1}}}
\nc{\dq}[1]{\frac{d\,^{#1}q}{(2\pi)^{#1}}}

\nc{\bfT}{{\bf T }}

\nc{\cA}{{\cal A}}
\nc{\cB}{{\cal B}}
\nc{\cD}{{\cal D}}
\nc{\cE}{{\cal E}}
\nc{\cG}{{\cal G}}
\nc{\cH}{{\cal H}}
\nc{\cL}{{\cal L}}
\nc{\cO}{{\cal O}}
\nc{\cT}{{\cal T}}
\nc{\cN}{{\cal N}}
\nc{\cR}{{\cal R}}

\nc{\rvac}[1]{|{\cal O}#1\rangle}
\nc{\lvac}[1]{\langle{\cal O}#1|}
\nc{\rvacb}[1]{|{\cal O}_\beta #1\rangle}
\nc{\lvacb}[1]{\langle{\cal O}_\beta #1 |}
\nc{\bb}{\bar{\beta}}
\nc{\bt}{\tilde{\beta}}
\nc{\ctH}{\tilde{\cal H}}
\nc{\chH}{\hat{\cal H}}
\nc{\al}{\alpha}
\nc{\g}{\gamma}
\nc{\Del}{\Delta}
\nc{\e}{\textrm{e}}
\nc{\eps}{\epsilon}
\nc{\lam}{\lambda}
\nc{\Om}{\Omega}
\nc{\ve}{\varepsilon}
\nc{\mn}{{\mu\nu}}
\nc{\vp}{\varphi}

\nc{\rf}[1]{(\ref{#1})}
\nc{\nn}{\nonumber \\*}
\nc{\bfB}{\bf{B}}
\nc{\bfv}{\bf{v}}
\nc{\bfx}{\bf{x}}
\nc{\bfy}{\bf{y}}
\nc{\vx}{\vec{x}}
\nc{\vy}{\vec{y}}
\nc{\oB}{\overline{B}}
\nc{\oI}{\overline{I}}
\nc{\oR}{\overline{R}}
\nc{\rar}{\rightarrow}
\nc{\ti}{\times}
\nc{\slsh}{\hskip-5pt/}
\nc{\sm}{Standard~Model~}
\nc{\MP}{M_{\rm Pl}}
\nc{\mpl}{M_{\rm Pl}}
\nc{\tp}{t_{\rm Pl}}

\nc{\pmin}{p_{\rm min}}
\nc{\pmax}{p_{\rm max}}
\nc{\fo}{f_0}
\nc{\foi}{f_{0,i}\,}
\nc{\fop}{f_0^P}
\nc{\fou}{f_0^U}

\nc{\eff}{{\rm eff}}
\nc{\MT}{M_{\rm T}}
\nc{\ML}{M_{\rm L}}
\nc{\kk}{\vek{k}}
\nc{\pp}{{\rm p}}
\nc{\pt}{\partial_t}
\nc{\half}{{1\over 2}}
\nc{\w}{\omega}
\nc{\uhat}{\hat{U}_\w}

\nc{\etal}{\mbox{\it et al.}}
\nc{\ie}{{\it i.e. }}
\nc{\eg}{{\it e.g. }}
\nc{\trh}{T_{\rm RH}}
\nc{\ad}{{a'\over a}}
\nc{\bd}{{b'\over b}}
\nc{\Rd}{{R'\over R}}
\nc{\diag}{{\textrm{diag}}}
\nc{\mato}[1]{\tilde{#1}}
\nc{\sinn}{\textrm{sinn}}
\nc{\sech}{\textrm{sech}}
\nc{\I}{\textrm{I}}
\nc{\II}{\textrm{II}}
\nc{\III}{\textrm{III}}
\nc{\vev}[1]{\langle #1 \rangle}
\nc{\hyp}{\,\; F_{1{\hskip -16pt}2}{\hskip 11pt}}
\nc{\brhom}{\overline{\rho}_M}
\nc{\brho}{\overline{\rho}}
\nc{\rhob}{\overline{\rho}}
\nc{\Pb}{\overline{P}}
\nc{\bH}{\overline{H}}
\nc{\ep}{{1+4\eps}}

\nc{\deriv}[2]{ 
\frac{\mathrm{d}#1}{\mathrm{d}#2}
}
\nc{\Mnu}{M_\nu}
\nc{\bee}{\begin{equation}}
\nc{\ene}{\end{equation}}
\nc{\hdp}{\sigma_8 (\Omega_{\rm m}/0.3)^{0.37}}
\nc{\avis}{\alpha_{vis}}
\nc{\cvis}{c^2_{vis}}
\nc{\clam}{c^2_{lam}}

\def\smiley{\hbox{\large$\bigcirc$\hspace{-.80em}%
\raise.2ex\hbox{$\cdot\cdot$}\kern-.61em    
\lower.2ex\hbox{\scriptsize$\smile$}}\ }

\def\frowney{\hbox{\large$\bigcirc$\hspace{-.80em}%
\raise.2ex\hbox{$\cdot\cdot$}\kern-.635em
\lower.2ex\hbox{\scriptsize$\frown$}}\ }

\begin{document}

\title{Bayesian analysis of an anisotropic universe model:
   systematics and polarization}
\author{Nicolaas E. Groeneboom \altaffilmark{1,2}, Lotty Ackerman\altaffilmark{3,4}, Ingunn Kathrine Wehus \altaffilmark{5,3} and Hans Kristian Eriksen\altaffilmark{1,2}}

\email{nicolaag@astro.uio.no}
\email{h.k.k.eriksen@astro.uio.no}
\email{lotty@theory.caltech.edu}
\email{i.k.wehus@fys.uio.no}

\altaffiltext{1}{Institute of Theoretical Astrophysics, University of
  Oslo, P.O.\ Box 1029 Blindern, N-0315 Oslo, Norway}

\altaffiltext{2}{Centre of Mathematics for Applications, University of
  Oslo, P.O.\ Box 1053 Blindern, N-0316 Oslo, Norway}

\altaffiltext{3}{California Institute of Technology, 1200 E. California blvd, Pasadena, CA 91125}

\altaffiltext{4}{Texas Cosmology Center, The University of Texas at Austin, TX 78712}

\altaffiltext{5}{Department of Physics, University of
  Oslo, P.O.\ Box 1048 Blindern, N-0316 Oslo, Norway}

\date{\today}

\begin{abstract} 

  We revisit the anisotropic universe model previously developed by
  Ackerman, Carroll and Wise (ACW), and generalize both the
  theoretical and computational framework to include polarization and
  various forms of systematic effects. We
  apply our new tools to simulated WMAP data
  in order to understand the potential impact of asymmetric
  beams, noise mis-estimation and potential Zodiacal light emission. We find that neither has any
  significant impact on the results. We next show that the previously
  reported ACW signal is also present in the 
  1-year WMAP temperature sky map presented by \cite{liu2009}, where data cuts are more
  aggressive. Finally, we reanalyze the 5-year
  WMAP data taking into account a previously neglected
  $(-i)^{l-l'}$-term in the signal covariance matrix. We still find a strong
  detection of a preferred direction in the temperature map. Including
  multipoles up to $\ell=400$, the anisotropy amplitude for the W-band
  is found to
  be $g = 0.29 \pm 0.031$, nonzero at $9 \sigma$. However, the corresponding
  preferred direction is also shifted very close to the ecliptic poles at
  $(l,b)= (96,30)$, in agreement with the analysis of
  \cite{hanson:2009}, indicating that the signal is aligned along the
  plane of the solar system. This strongly suggests that the signal is not of cosmological
  origin, but most likely is a product of an unknown systematic
  effect. Determining the nature of the systematic effect is of vital
  importance, as it might affect other cosmological conclusions from
  the WMAP experiment. Finally, we provide a forecast for the Planck
  experiment including polarization.

\end{abstract}

\keywords{cosmic microwave background --- cosmology: observations --- 
methods: numerical}

\maketitle

\section{Introduction}   
\label{sec:introduction}
In recent years, the study of the cosmic microwave background (CMB) has
proved to be the most fruitful addition to our understanding of the
early universe. Observations of the CMB anisotropies, like those
obtained by the Wilkinson Microwave Anisotropy Probe (WMAP) experiment
\citep{bennett:2003, hinshaw:2007}, have provided us with incomparable
insight on the composition of structure in our universe. Combined with
previous experimental knowledge and a sound theoretical framework, the
concordance model of $\Lambda$CDM has been established.

The $\Lambda$CDM model relies on the framework of inflation. Inflation
was initially proposed as a solution to the horizon and flatness
problem \citep{guth:1981}. Additionally, it established a highly
successful theory for the formation of primordial density
perturbations, providing the required seeds for the large-scale
structures (LSS).  Eventually, these later gave rise to the
temperature anisotropies in the cosmic microwave background radiation
that we observe today \citep{guth:1981,linde:1982, muhkanov:1981,
  starobinsky:1982, linde:1983, linde:1994, smoot:1992, ruhl:2003,
  runyan:2003, scott:2003}.

One of the predictions from inflation is that the observed universe
should be nearly isotropic on large scales. However, anomalies
found in the CMB during the recent years \citep{de Oliveira-Costa:2004, vielva:2004, eriksen:2004a} suggest that
anisotropic inflationary models should be
considered.  A specific example is the generalized model presented by
\citet{ackerman:2007}, which considers violation of rotational
invariance in the early universe. A general framework for describing
similar models was presented by \citet{pullen:2007}. 

\cite{himmel:2009a, himmel:2009b} showed that the anisotropic
inflationary background of the ACW model
characterized by a fixed-norm vector field ultimately is
unstable. However, the parametrization of the signal covariance  
matrix is independent of that unstable model, and is very useful.  
It represents general correlations induced by
rotations in the CMB at a phenomenological level. Several papers have
recently investigated the properties of the ACW model with extensions  
\citep{hou:2009,
karviauskas:2009, dimopoulos:2009, carroll:2008}.

Work in this field suggests that the 5-year WMAP data contains a significant ACW
anisotropic signal, corresponding to a 3.8$\sigma$ detection in the
W-band \citep{groeneboom:2008b}. A more recent paper by
\cite{hanson:2009} points out that the
direction is incorrect due to a neglected factor of $(-i)^{l-l'}$
corrected in a later version of \cite{ackerman:2007},
yielding an ACW-signal in which the preferred direction is located
very close to the ecliptic poles.

In this paper, we re-analyze the 5-year WMAP data including the
previously neglected $(-i)^{l-l'}$-factor, and investigate whether traces
of the ACW anisotropic contribution signal are still evident. The analysis will, as
previously, be performed with the CMB Gibbs sampling framework
\citep{jewell:2004, wandelt:2004, eriksen:2004b}, which by
\cite{groeneboom:2008b} was included to allow for non-diagonal, but
sparse covariance matrices. 
This framework allows for exact Bayesian analysis of high-resolution
CMB data with a
non-diagonal CMB signal covariance matrix. The isotropic method has already
been applied several times to the WMAP data \citep{odwyer:2004,
  eriksen:2007a, eriksen:2007b, eriksen:2008b}, and has already
been extended to take into account polarization
\citep{larson:2007} and internal component separation
\citep{eriksen:2008a}. 

In our
re-analysis of the 5-year WMAP temperature data, we confirm that the
direction is shifted to the ecliptic poles, at a greatly
increased significance. As the north and
south ecliptic poles are aligned with our solar system, the ACW signal in the WMAP data is therefore most likely
a systematic effect and not of cosmological origin. This is in
complete agreement with \cite{hanson:2009}.

It has not yet been possible to 
fully rule out whether any known systematic effect could have
contributed  to the signal. In theory, either asymmetric beams, mis-estimated noise
or even the Zodiacal light could have affected the detection of the
ACW-signal. In this paper, we consider these three effects and
conclude that neither have any effect on the ACW-signal.

Until now, the framework has only supported temperature-temperature
correlations (TT). Here, we extend the mechanics to include E-mode
correlations (EE), including cross-mode correlations
(TE). We then provide a forecast for the upcoming Planck experiment,
considering simulated T+E maps. The Planck data will hopefully be able to rule out all doubts about
the origin of the ACW signal.

\section{The ACW model with polarization}
\label{sec:extension}
We are interested in the signatures that the ACW model would leave on
the polarization of the CMB and focus our attention on the scalar
perturbations.  This calculation was first performed by \cite{pullen:2007}.
Observing the CMB sky in the direction $\hat{\bf e}$
provides information of the E-mode polarization constructed from the
Stokes parameters $Q(\hat{\bf e})$ and $U(\hat{\bf e})$, as well as
the temperature $T(\hat{\bf e})$.  One can express the respective maps
in terms of the spherical-harmonic coefficients $a_{E,lm}$ and
$a_{T,lm}$ which are given for each $X=\{E,T\}$ by
\begin{equation}
\small
a_{X,lm} = \int {\rm d} \Omega_{\bf e} Y_{lm}^{*}({\bf e}) \int{\rm d}
{\bf k} \,  \delta ({\bf k}) \left({2l+1 \over 4 \pi}\right) 
(-i)^l P_l({\hat {\bf k}}\cdot {\bf e})\Theta_{X,l}^{(S)}(k).
\label{alm}
\end{equation}
Here, $Y_{lm}({\bf e})$ denotes the spherical harmonics, $P_{l}({\bf
  k})$ are the Legendre polynomials, and $\Theta_{X,l}^{(S)}(k)$ is
the $l$th moment of the transfer function of scalar modes, for either
temperature or polarization.  Further, $\delta ({\bf k})$ is a random
variable that characterizes the initial amplitude of the mode and
satisfies
\begin{equation}
\langle \delta({\bf k})  \delta^*({\bf q})\rangle= P'({\bf k})\delta^3({\bf k}-{\bf q}).
\end{equation}
The ACW model proposes that if we drop the assumption of statistical
isotropy by having a preferred direction $\hat{\bf n}$ during
inflation, the primordial power spectrum at leading order has the form
\begin{equation}
P'({\bf k})=P(k)\left(1+g(k)({\hat {\bf k}}\cdot \hat{\bf n})^2  \right).
\label{powerspectrum}
\end{equation}
Here, $g(k)$ is a general function of $k$, which  ACW argue is
well approximated by a constant, $g_*$. 

To study the statistics of the CMB produced by the scalar
perturbations we need the power spectrum of the $T$, $E$ modes and the
cross-correlation between them.  Using the expressions (\ref{alm}) and
(\ref{powerspectrum}) we can write the various correlations for
$X=\{E,T\}$ as
\begin{equation}\label{aaa}
\langle a_{X,lm} a^*_{X',l'm'}  \rangle = \delta_{ll'}\delta_{mm'} \, C^{XX'}_{l,l}  +  g_* \, \xi_{lm;l'm'} \, C^{XX'}_{l,l'},
\end{equation}\
where the $C^{XX'}_{l,l'} $ are given by
\begin{equation}
C^{XX'}_{l,l'}=(-i)^{l-l'}\int_0^{\infty} {\rm d}k k^2 P(k)\Theta^{(S)}_{X,l}(k)\Theta^{(S)}_{X',l'}(k).
\label{bbb}
\end{equation}
The coefficients $\xi_{lm;l'm'}$ encode the departure from isotropy
and connect $l$ with $l' = \{l,l \pm 2\}$ and $m$ with $m'=\{m,m \pm
1, m\pm 2\}$ \citep{ackerman:2007}. Note that the factor of $(-i)^{l-l'}$ was missing in the first version of the paper.  

\section{The polarized anisotropic CMB Gibbs sampler}
\label{sec:gibbs}
CMB data observations can be modeled as:
\begin{equation}
  \mathbf d = \mathbf A \mathbf s + \mathbf n, 
\end{equation}
where $\mathbf{d}$ represents the observed data, $\mathbf{A}$ denotes
convolution by an instrumental beam, $\textbf s(\theta,\phi) =
\sum_{\ell,m} a_{\ell m} Y_{\ell m}(\theta,\phi)$ is the CMB sky
signal represented in either harmonic or real space and $\textbf n$
is instrumental noise. It is generally a good approximation to assume both the CMB and noise
to be zero mean Gaussian distributed variates, with covariance
matrices $\mathbf{S}$ and $\mathbf{N}$, respectively. In harmonic
space, the signal covariance matrix is defined by $\textbf S_{\ell
  m,\ell'm'} = \left< a_{\ell m} a_{\ell 'm'}^* \right>$. In the
isotropic case, this matrix is diagonal. The connection to cosmological parameters
$\omega$ is made through this covariance matrix.  Finally, for
experiments such as WMAP, the noise is often assumed uncorrelated
between pixels, $\textbf N_{ij} = \sigma_{i}^2 \delta_{ij}$, for
pixels $i$ and $j$, and noise RMS equals to $\sigma_{i}$.

Let $\omega$ denote a set of cosmological parameters. 
Our goal is to compute the full joint posterior $P(\omega |
\mathbf{d})$, which is given by $P(\omega
|\mathbf{d}) \propto P(\mathbf{d}| \omega ) P( \omega) = \mathcal{L}(
\omega) P(\omega),$ where $\mathcal{L}( \omega )$ is the likelihood
and $P(\omega)$ a prior. For a Gaussian data model, the likelihood
is expressed as:
\begin{equation}
  \mathcal L(\omega) \propto \frac{e^{-\frac{1}{2}\mathbf{d}^T
      \mathbf{C}^{-1}(\omega)\mathbf{d}}}{\sqrt{|\mathbf{C(\omega)}|}}.
\label{eq:likelihood}
\end{equation}
where $\mathbf{C}=\mathbf{S}+\mathbf{N}$ is the total covariance matrix.

\subsection{The Gibbs sampler}
The problem of extracting the cosmological signal $\mathbf s$  and
$\omega$ from the
full signal by Gibbs sampling was addressed by \citet{jewell:2004}, \citet{wandelt:2004}
and \citet{eriksen:2004b}. The CMB Gibbs sampler is an exact Monte
Carlo Markov chain (MCMC) method that assumes prior knowledge of the
conditional distributions in order to gain knowledge of the full joint
distribution. A significant fraction of the CMB data is completely
dominated by galactic foreground, and about $20\%$ of the data needs to
be removed. This might sound trivial, but in reality it complicates
processes as the spherical harmonics no longer are
orthogonal. The
Gibbs sampler solves this problem intrinsically, as the galaxy mask
becomes a part of the framework \citep{groeneboom:2009b}.  

The main motivation for introducing the CMB Gibbs sampler is
the drastically improvement in scaling. With conventional MCMC methods,
one needs to sample the angular power spectrum, $C_\ell= \langle
a_{\ell m} a_{\ell m}^* \rangle$, from the distribution  $P(C_\ell | \mathbf d)$,  
which scales as $\mathcal O(N_{\textrm{pix}}^3)$, where $N_{\textrm{pix}}$ is the size of the covariance
matrix. For a white noise case, the Gibbs
sampler reduces this to $\mathcal O(N_{\textrm{pix}}^{1.5})$. In other
words, the Gibbs sampler enables effective sampling in the high-$\ell$ regime.

\subsection{Sampling scheme}

In order to sample from the full joint distribution
$P(C_\ell, \omega, \mathbf s |\mathbf d)$ using the Gibbs sampler, we
must know the exact conditional distributions $P(\mathbf s
| C_\ell, \omega, \mathbf  d)$ and $P(C_\ell, \omega | \mathbf s )$. 
The Gibbs sampler then proceeds by alternating sampling from each of 
these distributions:  
\begin{align}
(C_\ell, \omega)^{i+1} \leftarrow& P(C_\ell, \omega| \mathbf{s}^i,
  \mathbf{d}) \\
\mathbf{s}^{i+1} \leftarrow& P(\mathbf{s} | (C_\ell, \omega)^{i+1}, \mathbf{d}).
\end{align}
The first conditional distribution is expressed as: 
\begin{equation}
  P(C_\ell, \omega | \mathbf s, \mathbf d) = \frac{e^{-\frac{1}{2}\mathbf s^T\mathbf
      S(\omega)^{-1}\mathbf s}}{\sqrt{|\mathbf S(\omega)|}},
\label{eq:invgamma}
\end{equation}
and is distributed according to
an inverse Gamma function with $2\ell+1$ degrees of freedom. The remaining
conditional distribution is
\begin{equation}
  P(\mathbf s | C_\ell, \omega, d) \propto e^{-\frac{1}{2}(\mathbf s-\hat{\mathbf{s}})^T(\mathbf S(\omega)^{-1} + \mathbf N^{-1})(\mathbf s-\hat{\mathbf{s}})},
\end{equation}
where $\hat{\mathbf{s}} = \mathbf N^{-1}\mathbf d$. In other words,
$P(\mathbf s | C_\ell, \omega, \mathbf d)$ is a
Gaussian distribution with mean $\hat{\mathbf{s}}$ and covariance $(\mathbf S(\omega)^{-1} +
\mathbf N^{-1})^{-1}$. Numerical methods for sampling from these
distributions were discussed by \cite{groeneboom:2009b}, and the details on
how the polarization covariance  matrix was numerically implemented can be found in Appendix \ref{sec:covariancematrix}.

\section{Re-analysis of 5-year temperature WMAP data}
\label{sec:correction}

\begin{figure*}
\mbox{\epsfig{file=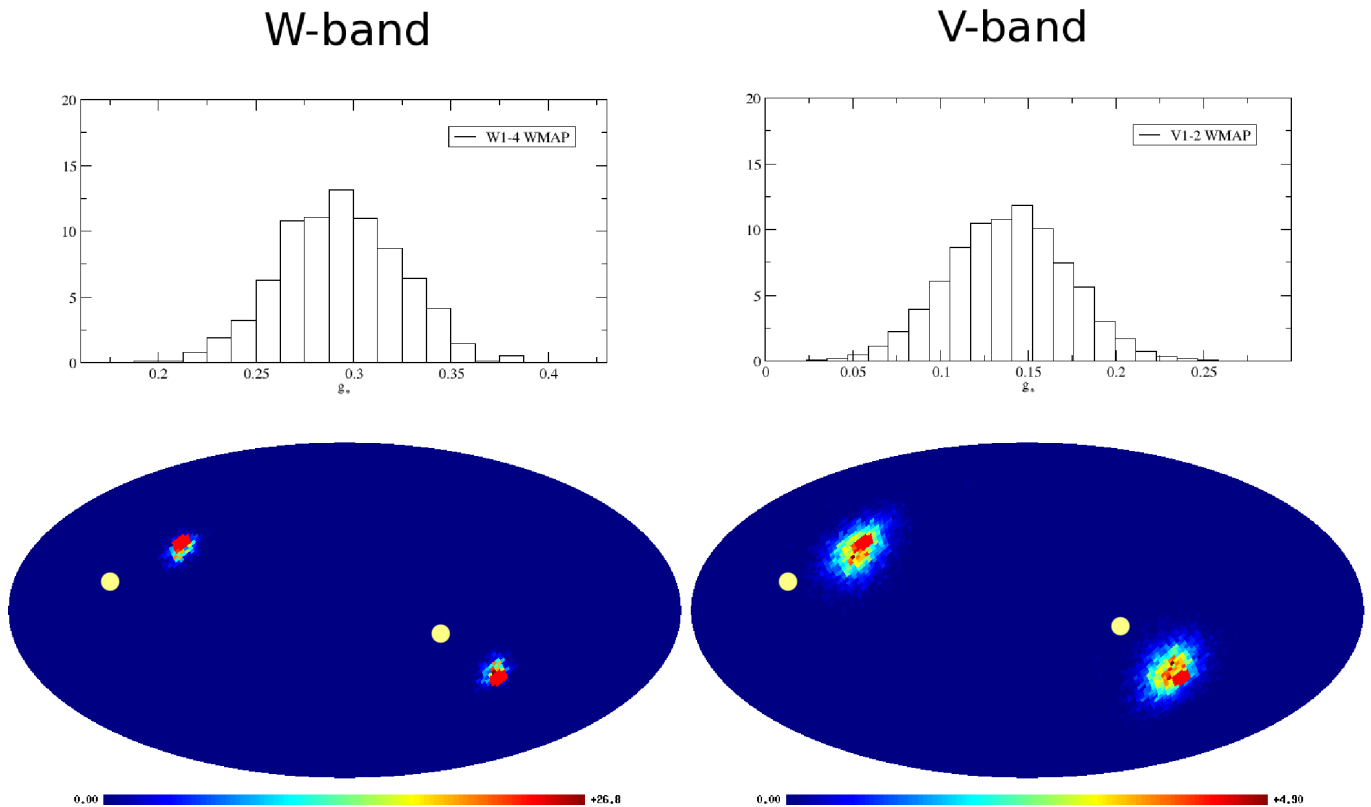, width=\linewidth,clip=}}
\caption{W and V-band posteriors for the temperature analysis, using
  $\ell_{\textrm{cutoff}}=400$ and the KQ85 mask. The north and south
  ecliptic poles are marked with a red circle. Note how the posterior peaks
  correspond with the ecliptic poles. The yellow circles indicate the
  direction from the previous analysis by \cite{groeneboom:2008b}.
}
\label{fig:res_bands}
\end{figure*}
\cite{ackerman:2007} and \cite{hanson:2009} pointed out an error
in the expression for the off-diagonal covariance matrix. 
The expression for the signal covariance matrix (\ref{aaa}-\ref{bbb})
now includes a previously neglected factor of $(-i)^{l-l'}$. 
For the ACW-covariance matrix that correlate scales with $\ell = \ell' \pm 2$,
the only difference in contribution is $(-i)^{\pm 2}=-1$, negating
the off-diagonal terms. 
\cite{hanson:2009} claims that the
ACW-signal direction in the 5-year WMAP data is located  at the
ecliptic poles, and not at $(l,b) = (110^\circ, 30^\circ)$, as
presented by \cite{groeneboom:2008b}. In light of the new results, we
perform a new full temperature analysis of the WMAP data and
investigate whether the neglected factor has any impact on the
resulting posteriors.

\label{sec:wmap5}
\subsection{Data}
We consider the five-year WMAP temperature sky maps
\citep{hinshaw:2009}, and analyze the Q-, V and W-bands (41, 61 and 94 
GHz), where the W and V bands are assumed to be the cleanest WMAP bands in terms of residual
foregrounds. We adopt the template-corrected, foreground reduced maps
recommended by the WMAP team for cosmological analysis, and impose
the KQ85 masks \citep{gold:2008}, which remove 18\% of the sky. 
Point source cuts are imposed in both masks.

We analyze the data frequency-by-frequency, and consider the
combinations V1+V2, Q1+Q2 and W1 through W4. The noise RMS patterns
and beam profiles are taken into account for each difference assembly
map (DA) individually. The noise is assumed uncorrelated.  All data
used in this analysis are available from LAMBDA
\footnote{http://lambda.gsfc.nasa.gov/}.

\subsection{Results}

\begin{deluxetable}{lcccc}
\tablewidth{0pt}
\tablecaption{Summary of marginal posteriors from WMAP5  \label{tab:distributions}} 
\tablecomments{The values
  for $g_*$ indicate posterior mean and standard deviation. The
  ecliptic poles are located at $\pm (96^\circ, 30^\circ).$} 
\tablecolumns{4}
\tablehead{ Band & $\ell$ range  & Mask & Amplitude $g_*$ & Direction $(l,b)$ }
\startdata
W1-4 & $2-400$   &    KQ85    &   $0.29 \pm 0.031$               &
$(94^{\circ}, 26^{\circ}) \pm 4^{\circ} $                       \\ 
V1-2 & $2-400$   &    KQ85    &   $0.14 \pm 0.034$               &
$(97^{\circ}, 27^{\circ}) \pm 9^{\circ} $                       \\ 
Q1-2 & $2-300$   &    KQ85    &   $-0.18 \pm 0.040$               & 
$(99^{\circ}, 28^{\circ}) \pm 10^{\circ} $ 
\enddata
\label{tab:WMAP}
\end{deluxetable}

The results from our analysis are presented
in Table \ref{tab:WMAP}, and the posteriors are shown in Figure
\ref{fig:res_bands}. The strongest detection is still present in the
W-band, where $g_* = 0.29 \pm 0.031$, corresponding to a $9 \sigma$ detection.
However, the correction term mentioned above clearly has a significant
effect on the signal described by \cite{groeneboom:2008b}. 
The direction and the significance of the detection is
altered: For both the W-band and V-band analyses, the
preferred direction is now located at $(l,b) = \pm(96^\circ,30^\circ)$, very close to
the north/south ecliptic poles. In addition, the significance of the
signal in the W-band is increased from previously $3.8\sigma$ to
about $9\sigma$, showing that the neglected correction-term has
``forced'' the signal away from its true direction  - the north and
south ecliptic poles. The probability that this direction is a pure
coincidence is minimal, and the observed signal is therefore most
likely a product of systematics.  Another interesting fact is that the signal seems
to be frequency dependent, with a stronger signal in the W
bands than in the V bands. Further, the Q-bands seems to exhibit a
negative $g_*$, which suggests frequency dependence.

\section{Analysis of systematic effects}
\begin{figure*}
\mbox{\epsfig{file=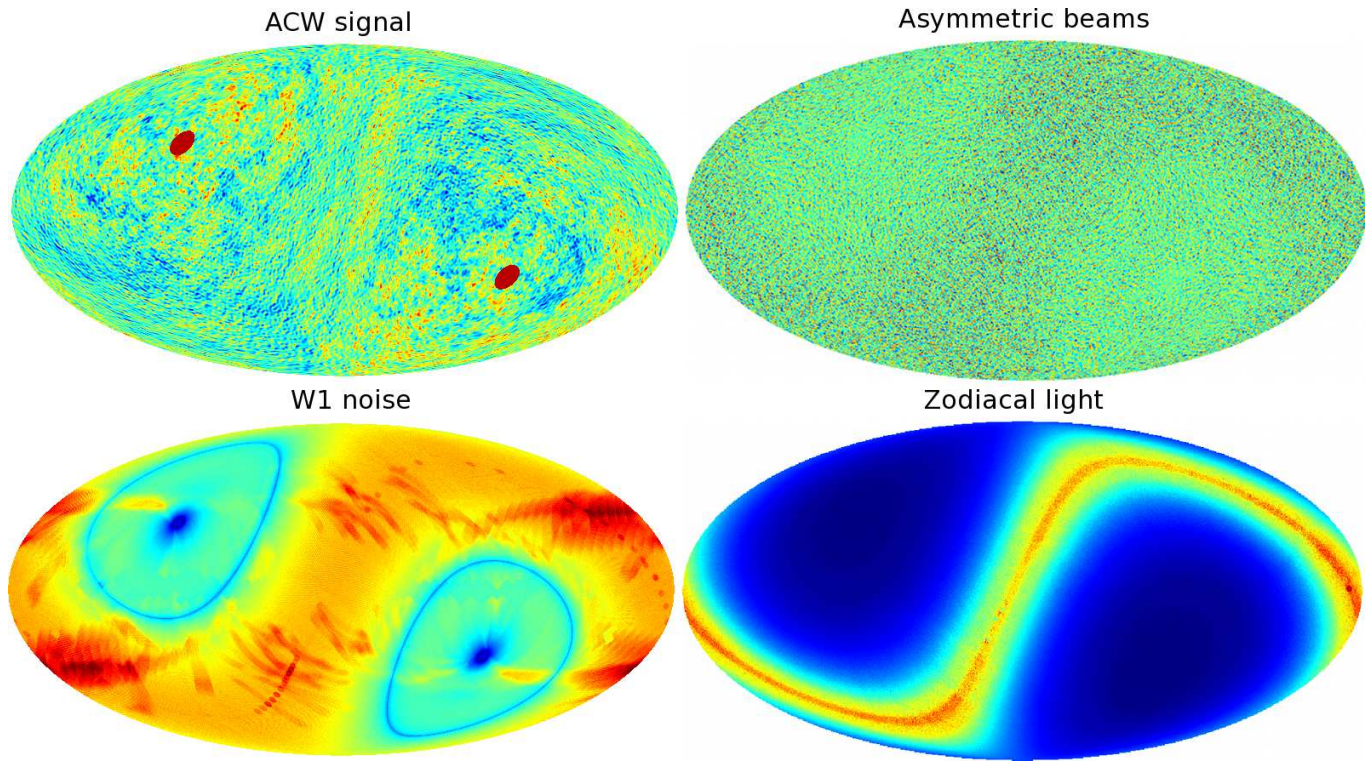, width=\linewidth,clip=}}
\caption{Various systematic effects compared with the ACW signal
  (upper left). Asymmetric beams (upper right), noise maps (bottom
  left) and the Zodiacal light template (lower right) are similar in
  shape to the ACW signal, and could therefore be thought to contribute
  to the  ACW signal in the WMAP data. 
}
\label{fig:systematics}
\end{figure*}

Before the
correction was introduced, we performed several tests on the
independent WMAP 5-year DA bands showing that the direction is both
existent and stable in bands.  The significance
was also slightly increased, and we are able to cover the signal up to
$\ell_{\textrm{max}} = 700$. We now proceed by investigating
various systematic effects as candidates for the observed signal. A visualization of
some possible sources of systematic effects together with a realization of the
ACW signal for comparison is presented in figure
\ref{fig:systematics}; Asymmetric beams (upper right), noise RMS maps (bottom
left) and the zodiacal light template (lower right).

\subsection{Impact of noise mis-estimation}
One of the possible candidates for generating
the ACW-signal found by
\cite{groeneboom:2008b} is noise mis-characterization. 
Previous work done by \citep{groeneboom:2008b} showed that correlated noise levels have little or no
effect on the signal. However, it might be possible that noise with
incorrect RMS specifications could give rise 
to a signal similar as the ACW signal. We therefore
perform one more analysis to test noise sensitivity.

\cite{groeneboom:2009a} discovered that the noise levels
provided by the WMAP team were slightly off by about $0.5 -
1$ \%. While this error is small enough to not significantly affect most
cosmological analyses, it is conceivable that incorrect noise levels
could contribute to a signal similar to the ACW-model.

We therefore simulate a V1 map with $5$ \% incorrect V1 noise, i.e.,
the noise is multiplied with $1.05$ before it is added to the map. The
analysis is done with the $KQ85$ mask. The $\chi^2$ comes out about
$ 6$\% above the expected value, recording that the incorrect noise
is measured by the Gibbs sampler. However, the posteriors still show
a zero-detection of the ACW-model, with an anisotropic amplitude of 
$g_* = 0.01 \pm 0.05$.  This indicates that incorrect
noise levels have little or no effect on the ACW-signal.

\subsection{Impact of asymmetric beams}
Another issue with the analysis of \cite{groeneboom:2008b} is
whether the asymmetric beams of the WMAP detectors could have given
rise to a signal similar to the ACW model. \cite{ingunn:2009}
established a full framework for simulating WMAP maps 
with asymmetric beams. An example of contribution from  asymmetric beams
on WMAP maps is presented in Figure \ref{fig:systematics}.
The authors also provided a set of 10
simulated maps with asymmetric beams. We now perform a 
Bayesian analysis on these maps, together with an analysis on isotropic
simulated maps with symmetric beams for comparison.  
\begin{figure}
\mbox{\epsfig{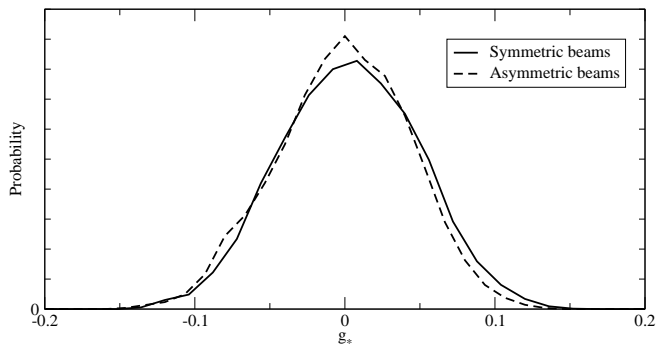}}
\caption{An analysis of the same isotropic map convolved with
  symmetric (black)  and asymmetric (red) beams. Note how both results
are consistent with $g_*=0$}
\label{fig:res_asymm2}
\end{figure}

The test data are set up as such: We simulate isotropic test maps
with the best-fit $\Lambda$CDM power spectrum, and convolve them with 
the standard symmetric V-beams. We
then add  V-band noise RMS to the maps, and 
analyze the test maps. We then perform the same analysis on the
V-band maps from \cite{ingunn:2009}, which were produced with
asymmetric beams. Both analyses are
done using multipoles $l_{\textrm{max}}=700$ and
$l_{\textrm{max}}^{\textrm{cutoff}}=512$, with a standard  V-band
setup and the KQ85 mask. 

The
posteriors for the anisotropy amplitude $g_*$ are shown in
Figure \ref{fig:res_asymm2}, with both having 
$g_* = -0.01 \pm 0.05$. It should be clear that
asymmetric beams do not produce effects in the CMB similar to the
ACW-model, as the analysis show no trace of any signal detection. 

\subsection{Zodiacal light }
In this paper, we have seen that the ACW signal in the WMAP data has
shifted to the ecliptic poles. This indicates that the signal is most
likely not of cosmological origin, as it is strongly aligned in the plane of the
solar system. An interesting question is whether the ACW-signal is
connected to the Zodiacal light. Zodiacal light is produced by
sun-rays reflecting off dust particles sharing the same orbit as the
earth, and share a similar overall structure as the ACW-signal. An
illustration of a zodiacal light template is presented in Figure
\ref{fig:systematics}, together with the estimated ACW signal in the
direction of the ecliptic poles. We perform three analyses of
realistic V-band simulations, where we co-add the Zodiacal light
template to simulated, isotropic V-band simulated maps. In the first run,
we add the template as it is, in the second and third analyses we
multiply the template with a factor of 10 and 100, respectively. In
all of the analyses, the posteriors resulted in zero-detections with
$g_*=0.0 \pm 0.045$ and no significant directions on the sky, with uniform
distributions. We therefore conclude that the Zodiacal light does not
have a significant contribution to the ACW signal in the WMAP data.

\subsection{Analyzing alternative WMAP data}
\cite{liu2009, liu:2009b} have developed an alternative framework for
building 1-year WMAP maps from raw data. The authors imposed
stronger constraints on data selection, removing almost $20 \%$ of the
time-ordered data. For instance, data for which the beam boresight distance from the planets 
are less than $7^\circ$ are removed, corresponding to the the antennae
main beam radius. The temperature map published by the WMAP team used a cut of 
only $1.5^\circ$ \citep{limon:2008}. \cite{liu2009}  also used an extended
KQ85-mask which removes $28.3 \%$ of the sky. \cite{liu:2009b}
claim that the pixels in the WMAP scan ring of a 
hot pixel are systematically cooled, where the strongest anti-correlations 
between temperatures of a hot pixel and its scan-ring appear at a 
separation angle of about $141^\circ$.  Due to the anti-correlation of pixels and the strict data cuts,
the temperature power spectrum obtained by \cite{liu2009} is
decreased on average by about $13 \%$, causing 
the best-fit cosmological parameters to change considerably. 

In order to see whether the anti-correlated pixels in the WMAP stream could have
contributed to the ACW signal in the WMAP data, we perform a full
temperature analysis on both the alternative temperature map provided
by \cite{liu2009} and
the original 1-year WMAP temperature map. The map used in our analysis is the
V1-band. The RMS noise map for the
alternative analysis is provided by \cite{liu2009}, while
the maps for the standard WMAP analysis were downloaded from the
Lambda site. The V1 beam is the same in both cases, as is the
extended KQ85 mask from \cite{liu2009}. If the ACW signal is detected
in the WMAP data but not the data from \cite{liu2009}, it might be an
indication that the WMAP team have included data that should have been
left out,
giving rise to a correlation structure similar to that of the ACW signal.

Analyzing the maps up to $\ell_{\textrm{max}} = 400$, we find that
both maps do contain a significant anisotropic signal, with $g_*
\sim 0.15 \pm 0.10$. This implies that the ACW-signal is
most likely a more intrinsic part of the WMAP data, and not due to the possible
anti-correlation of pixels.

\section{Forecasts for Planck with polarization}
The Planck satellite will provide us with high-resolution CMB data
of superior quality compared to previous CMB
experiments. The Planck experiment also provides high-resolution polarization
data, with an $\ell$-range up to 2500. As the Planck data are
independent from WMAP data, it will be very interesting to see
whether the ACW-signal is evident or not in the data. We therefore need to
investigate some anisotropic properties of typical Planck-data in
order to know what to expect and not expect. 

In this section, we set up a high-$\ell$ temperature analysis with
$\ell_{\textrm{max}}^{\textrm{cutoff}}=800$ and a joint temperature and polarization analysis
with $\ell_{\textrm{max}}^{\textrm{cutoff}}=400$. We then analyze the
maps to obtain the posterior means and standard deviation. We
continue by forecasting how the standard deviation of the anisotropic
amplitude posteriors should vary with multipoles $\ell$, as
done by \cite{groeneboom:2008b}.

\subsection{Validation of the polarized sampler}
\begin{figure}
\mbox{\epsfig{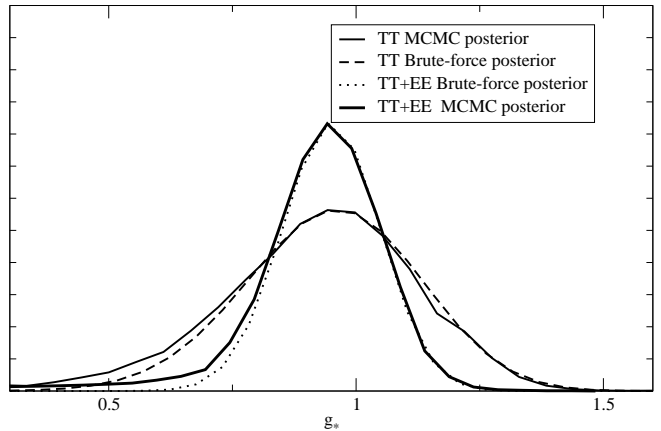}}
\caption{$g_*$ posteriors for several analysis of noiseless simulated $\ell_{\textrm{max}}=64$
  map, using both MCMC and brute-force calculations. Note how the
  polarization data narrows the distribution. 
}
\label{fig:simulated_posteriors}
\end{figure}

\begin{figure}
\mbox{\epsfig{file=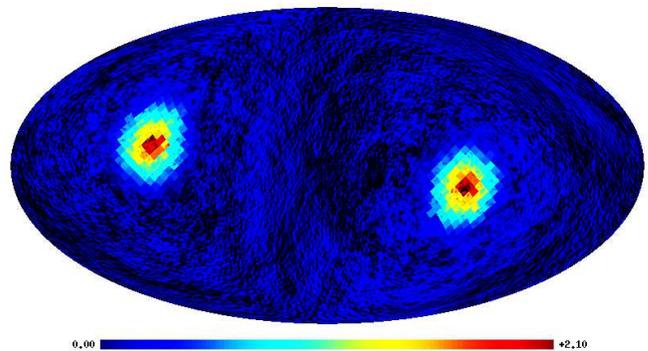, width=\linewidth,clip=}}
\caption{Posterior from a simulated set with $g_*=1.0$. The original
temperature ACW-signal in the input map can be seen in the background. Note
how the estimated direction corresponds well with the posterior.}
\label{fig:res_asymmetric}
\end{figure}

Before performing a full-scale analysis of simulated polarized Planck data, we
wish to validate our code. We therefore simulate a low-resolution $N_{\textrm{side}}=32$ map with E-
mode data included. Assuming an anisotropic amplitude of $g_*=1.0$, we perform both a brute-force
and a metropolis-hastings analysis of a full-sky map with no beam nor
noise. The resulting posteriors for the TT-case and the
TT+TE+EE-case are shown in Figure \ref{fig:simulated_posteriors}. It
is worth to note that the posterior is more narrow when including
polarization data, as there is more data available. A typical
posterior of the estimated direction $n$ together with the input TT+EE ACW-signal
is seen in Figure \ref{fig:res_asymmetric}. 

\subsection{Simulations}
\begin{deluxetable*}{lccccc}
\tablewidth{0pt}
\tablecaption{Summary of marginal posteriors from simulated Planck data  \label{tab:distributions}} 
\tablecomments{The values
  for $g_*$ indicate the posterior mean and standard deviation.} 
\tablecolumns{4}
\tablehead{ Simulated data & Input amplitude & $\ell$ range  & Mask & Estimated $g_*$  }
\startdata
Low-$\ell$ TT & $0.10$ &$2-400$   &    KQ85    &   $0.11 \pm 0.025$   \\ 
High-$\ell$ TT & $0.10$ &$2-800$   &    KQ85    &   $0.11 \pm 0.020$   \\ 
Low -$\ell$ TT+TE+EE & $0.10$ &$2-400$   &    KQ85    &   $0.10 \pm 0.020$  
\enddata
\label{tab:PLANCK}
\end{deluxetable*}
We now consider a Planck simulation.  We first simulate a
temperature-only ACW-anisotropic map with $n_{\textrm{side}}=1024$,
$\ell_{\textrm{max}}=2000$ and $\ell_{\textrm{cutoff}}=1024$, with a
preferred direction pointing towards $(\theta, \phi) = (57^{\circ},
57^{\circ})$ and an anisotropy amplitude of $g_* = 0.1$, using the
best-fit 5-year WMAP $\Lambda$CDM power spectrum
\citep{komatsu:2009}. The map is convolved with a Gaussian beam
corresponding to the $143$ GHz Planck channel, and white, uniform
noise is finally added. The beam FWHM for this frequency channel is
$7.1'$, and the temperature noise RMS per $N_{\textrm{side}}=1024$
pixel is $\sigma_T = 12.2\mu\textrm{K}$. The polarization noise RMS is
$\sigma_P = 23.3\mu\textrm{K}$ \citep{planck}.

\subsection{Results}

We perform three analyses of the simulated Planck sky map. The first
is an analysis on low-$\ell$
($l_{\textrm{max}}^{\textrm{cutoff}}=400$) temperature data, the
second high-$\ell$ ($l_{\textrm{max}}^{\textrm{cutoff}}=800$)
temperature data while the third is a low-$\ell$ analysis of TT+TE+EE
polarization data. The results are shown in Table \ref{tab:PLANCK},
where we reproduced the input parameters with typically $g_* = 0.11
\pm 0.025$. Note how the standard deviation of the posterior is lower
than for the WMAP case. This is to be expected, as higher multipoles
$\ell$ contribute more to the anisotropic effect, but not
significantly. This is due to the fact that the off-diagonal
correlation terms in the covariance matrix have a lower values on
smaller scales.

\begin{figure}
\mbox{\epsfig{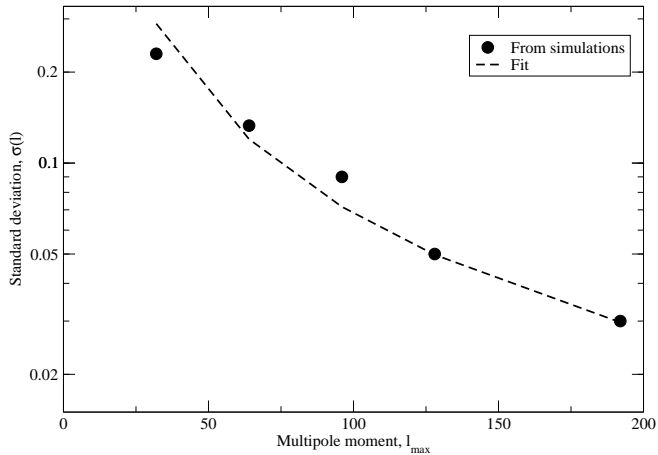}}
\caption{Estimated uncertainty in $g_*$ as a function of
  $\ell$ (black dots) and a best-fit power law
  function (red line) for cosmic variance limited data.}
\label{fig:planck_sigma}
\end{figure}

We  determine the standard deviation of the $g_*$
posterior as a function of multipoles $\ell$ by simulating a
unconvolved, noiseless isotropic
map including polarization data using the best-fit $\Lambda$CDM power
spectrum. We then analyze this map
for various $\ell$, obtaining the posterior distribution for each
run. The results are seen in Figure \ref{fig:planck_sigma}. Here, we see that $\sigma(\ell_*)$
is very close to a power law in $\ell$, in good
agreement with the arguments given by \citet{pullen:2007} and \cite{groeneboom:2008b}. The
best-fit power law function is
\begin{equation}
\sigma(\ell_{\textrm{high}}; g_*) = 0.0117\left(\frac{\ell_{\textrm{high}}}{400}\right)^{-1.27},
\end{equation}
and this can be used to produce rough forecasts for 
the Planck experiment including polarization. For instance, 
if both temperature and E-mode polarization data are available up to $\ell=512$,
then the standard deviation of $g_*$ is $\sigma(512) \sim 0.001$. This
is generally a factor two better than using temperature alone.

\section{Conclusions}   
\label{sec:conclusion}
We have generalized a previously developed Bayesian framework
to allow for exact analysis of any general anisotropic universe models that
predicts a sparse signal harmonic space covariance matrix, including
polarization data. This
generalization involved incorporation of a sparse matrix library into
the existing Gibbs sampling code called ``Commander''. 
We implemented support for this model in our codes, before demonstrating
and validating the new tools with appropriate simulations including
polarization data. First, we
compared the results from the Gibbs sampler with brute-force
likelihood evaluations, and then verified that the input parameters
were faithfully reproduced in realistic WMAP simulations.

We then considered a special case of anisotropic universe models,
namely the \citet{ackerman:2007} model which generalizes the
primordial power spectrum $P(k)$ to include a dependence on direction,
$P(\mathbf{k})$. The equations were however not complete, and the
analysis performed by \cite{groeneboom:2008b} has been re-done
including the previously neglected $(-i)^{l-l'}$-term. 

We then analyzed the five-year WMAP temperature sky maps, and
presented the updated WMAP posteriors of the ACW model. The
results from this analysis are in accordance with the results from
\cite{hanson:2009}, showing that the preferred direction is now
located at the ecliptic poles. This suggests that the signal
is most likely not of cosmological origin, and its origin must be
either from within the solar system or systematics. 

We have investigated four cases of systematic effects that
share similar structures with the ACW signal. We have shown
that neither asymmetric beams, the Zodiacal light, 
noise RMS mis-estimation nor possible pixel anti-correlations in the
WMAP data could have given rise to the observed signal. 

To summarize, we have shown that there exist a strong anisotropic
signal corresponding to the ACW signal in all the WMAP data that is aligned with the north and south
ecliptic poles. The probability that the axis should correspond so
closely to the ecliptic poles is very low, indicating that the signal
is due to a systematic effect. The signal makes up more than $5 \%$ of the
total power of the temperature fluctuations in the CMB. We have excluded
some of the possible candidates as source of the ACW signal.   
Determining the nature of the systematic effect will be of vital
importance, as it might affect other cosmological conclusions from
the WMAP experiment, and the upcoming Planck data will clearly be
invaluable for understanding the nature of this feature.

\begin{acknowledgements}
  We thank Liu Hao and Ti-Pei Li for supplying us with their 1-year WMAP
  data. We acknowledge use of the
  HEALPix \footnote{http://healpix.jpl.nasa.gov} software
  \citep{gorski:2005} and analysis package for deriving the results in
  this paper. We acknowledge the use of the Legacy Archive for
  Microwave Background Data Analysis (LAMBDA). Support for LAMBDA is
  provided by the NASA Office of Space Science. The authors
  acknowledge financial support from the Research Council of Norway.
\end{acknowledgements}

\appendix

\section{The covariance matrix}
\label{sec:covariancematrix}
\begin{figure}
\center
\mbox{\epsfig{file=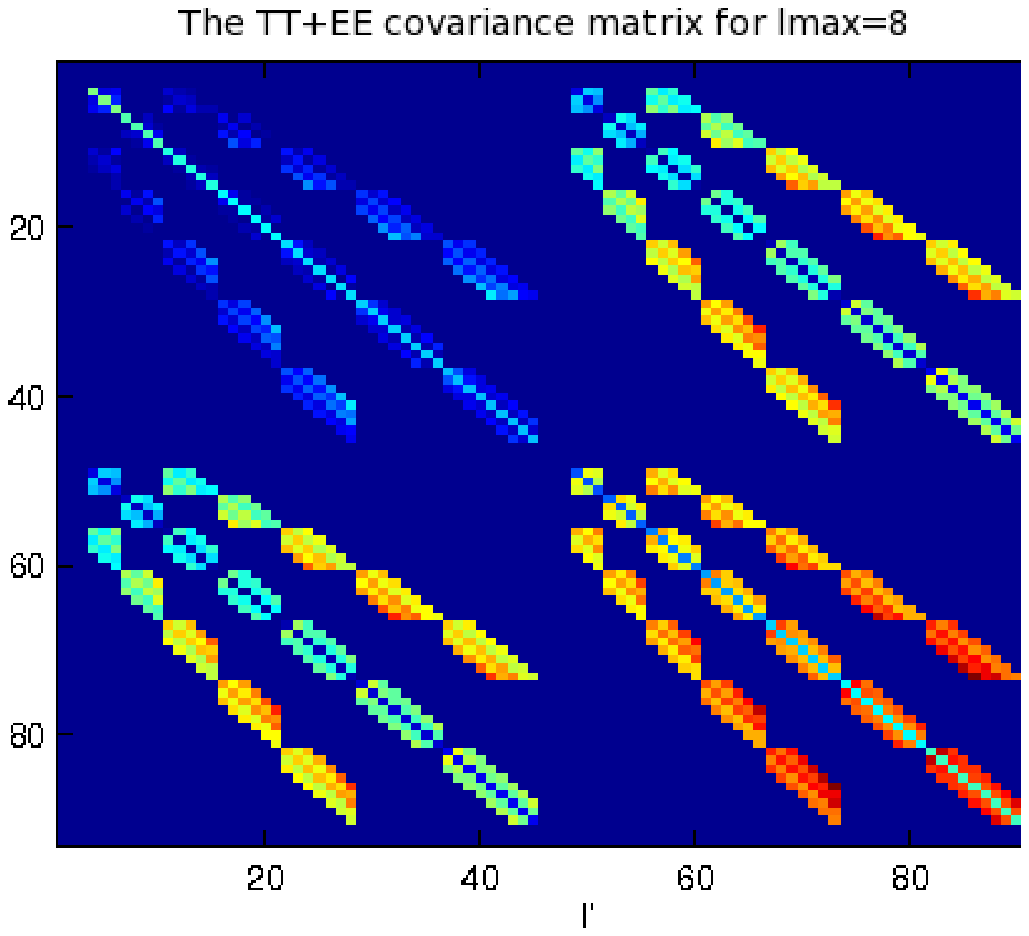, width=60mm,clip=}}
\mbox{\epsfig{file=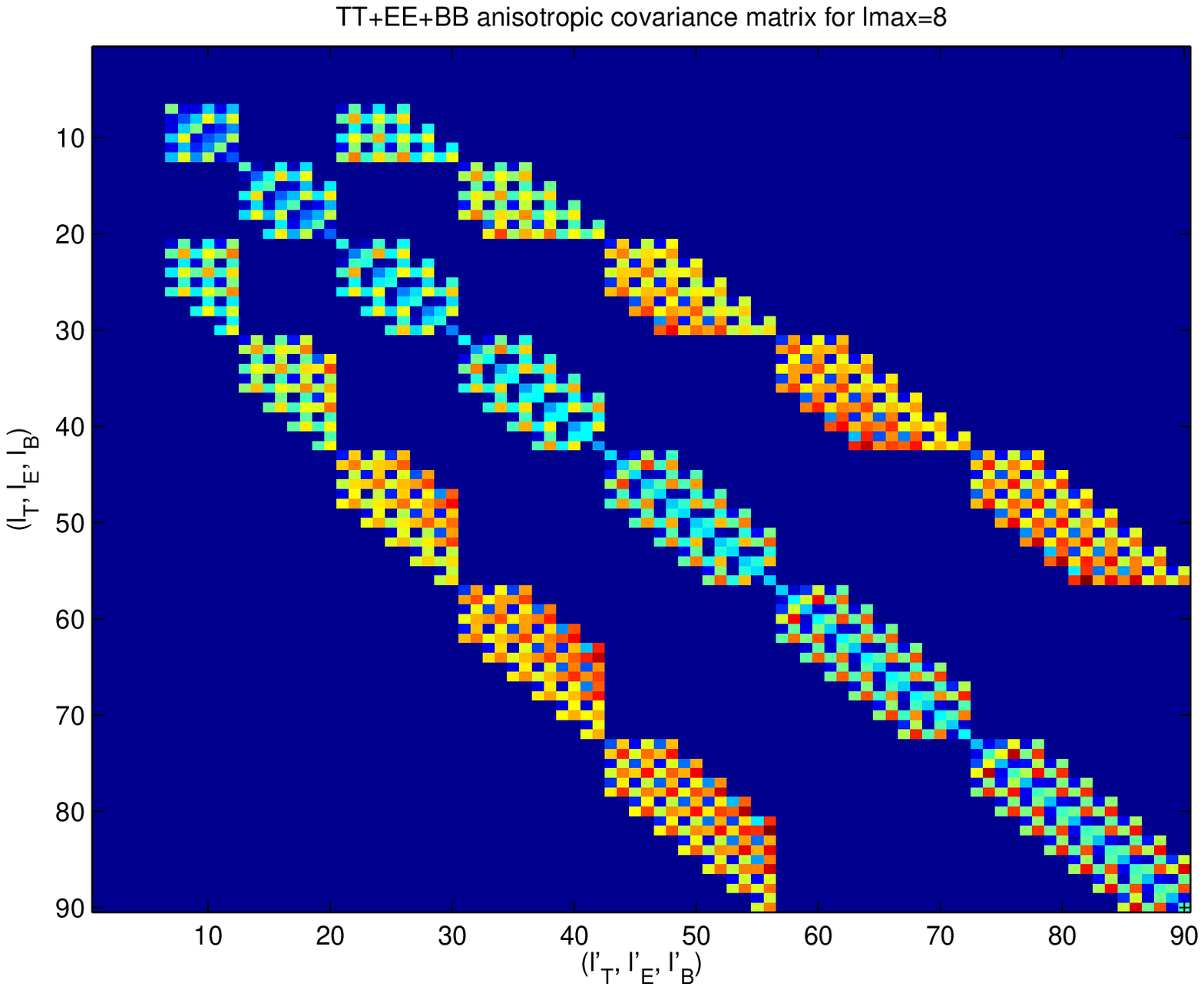, width=74mm,clip=}}
\caption{The ACW TT-EE covariance matrix (left) in the representation of
  Equation \ref{eq:covariancematrix1}. Diagonalizing this matrix
  turns out to be a nearly impossible task, forcing us to use another
  representation. The ACW TT-EE covariance matrix (right) in the representation of
  Equation \ref{eq:covariancematrix2}. Diagonalizing this matrix
  is similar to diagonalizing the TT-only ACW covariance matrix, and
  is more efficient.
}
\label{fig:covmatold}
\end{figure}

Even though we do not employ B-mode polarization data in the analysis
performed in this paper, the numerical framework still supports B-mode
polarization. In this section, we therefore describe the full TT+EE+BB
covariance matrix including correlations. In the previous analysis, only temperature-temperature anisotropic
correlations were considered. We now extend the framework to include
polarization, such that the Fourier coefficients become: 
\begin{equation}
\label{eq:alms}
a_{\ell m} = (a_{\ell m}^{TT}, a_{\ell
  m}^{EE}, a_{\ell m}^{BB}). 
\end{equation}
The covariance matrix $C_{\ell m, \ell'
  m'}$ can be expressed as:
\begin{equation}
\label{eq:covariancematrix1}
C_{\ell m, \ell' m'}=
\begin{pmatrix}
TT & TE & TB \\
TE & EE & EB \\
TB & EB & BB \\
 \end{pmatrix}.
\end{equation}
The existing framework for sampling anisotropic universe models in
FORTRAN was then altered to allow for polarization data, and whether
polarization is used is flagged through a parameter file. 
The off-diagonal TT+EE+BB anisotropic covariance matrix is presented in figure
\ref{fig:covmatold}. Note that the BB component is zero in this
plot. However, this straight-forward representation of the full covariance matrix is
too naive: performing a Cholesky-factorization (diagonalizing) of this matrix
for high $\ell$s are nearly impossible. Diagonalizing a matrix is more
efficient when
off-diagonal elements are close to the diagonal. However, the
(TT,EE,BB) representation of the matrix in figure
\ref{fig:covmatold} gives rise to elements spread around the full matrix. Typically,
Cholesky-factorization for such a TT-EE-BB representation breaks down for
$l_{\textrm{max}}=64$ due to the dense structure of the
upper-triangular decomposed L-matrix. 

To overcome this problem, we operate with a different representation
of the (TT,EE,BB)-matrix. Instead of building the matrix as
presented in equation \ref{eq:covariancematrix1}, we choose a
different way of expressing the matrix:
\begin{equation}
\label{eq:covariancematrix2}
C_{\ell m, \ell' m'}=
\begin{pmatrix}
\small
TT_{00} & TE_{00} & TB_{00} & TT_{10} &  \dots & EE_{n0}\\
TE_{00} & EE_{00} & EB_{00} & TE_{10} &  \dots & EB_{n0}\\
TB_{00} & EB_{00} & BB_{00} & TB_{10} &  \dots & BB_{n0}\\
TT_{01} & TE_{01} & TB_{01} & TT_{11} &  \dots & EE_{n1}\\
\vdots & \vdots  & \vdots & \vdots & \dots & \vdots\\
TB_{0n} & EB_{0n} & BB_{0n} & TB_{1n} & \dots & BB_{nn}\\
 \end{pmatrix}
\end{equation}
with corresponding $a_{\ell m}$s
\begin{equation}
  a_{\ell m} = (a^{T}_{00}, a^{E}_{00}, a^{B}_{00}, a^{T}_{01}, \dots,
  a^B_{l_{max}, m(l_{max})})
\end{equation}
As the EE and BB correlations share the same structure as the
stand-alone TT, the
complete covariance matrix will in this representation resemble the
original three-banded covariance matrix. The
elements are now much closer to the diagonal, solving the
problem of inefficient diagonalizing. The matrix representation is
depicted in figure \ref{fig:covmatold}.
Note that this representation is only used when multiplying the
matrices with vectors and performing Cholesky decompositions. Withing
the rest of the framework, the $a_{\ell m}$s are treated as in
equation \ref{eq:alms}.

\end{document}